\documentclass[12pt]{article}
\usepackage{natbib}
\usepackage{latexsym,makeidx,graphicx,subfig,rotate}
\usepackage[centertags]{amsmath}
\usepackage[active]{srcltx}
\usepackage{amsfonts}
\usepackage{amssymb}
\usepackage{amsthm}
\usepackage{newlfont}
\usepackage{amsfonts}
\usepackage{amssymb}
\usepackage{indentfirst}
\usepackage{graphics}
\usepackage{amsmath}
\allowdisplaybreaks

\usepackage{pdflscape} %PM: Package for landscape plots

\newfont{\Sc}{eusm10}

\newcommand{\cip}{\mbox{$\perp\!\!\!\perp$}}
\newcommand{\cb}{ \begin{eqnarray} }
\newcommand{\ce}{ \end{eqnarray} }

\usepackage{setspace}

\begin{document}

\title{Orthogonal prediction of counterfactual outcomes}
\author{Stijn Vansteelandt$^1$ and Pawe{\l} Morzywo{\l}ek$^{1,2}$
\bigskip \\
$^1$Ghent University \\
$^2$University of Washington}
\date{\today}

\maketitle
\doublespacing

Orthogonal meta-learners, such as DR-learner \citep{kennedy2020towards}, R-learner \citep{nie2021quasi} and IF-learner \citep{curth2020estimating}, 
are increasingly used to estimate conditional average treatment effects. They improve convergence rates relative to na\"{\i}ve meta-learners (e.g., T-, S- and X-learner \citep{kunzel2019metalearners}) through de-biasing procedures that involve applying standard learners to specifically transformed outcome data. This leads them to disregard the possibly constrained outcome space, which can be particularly problematic for dichotomous outcomes: these typically get transformed to values that are no longer constrained to the unit interval, making it difficult for standard learners to guarantee predictions within the unit interval.
To address this, we construct orthogonal meta-learners for the prediction of counterfactual outcomes which respect the outcome space. As such, the obtained i-learner or imputation-learner is more generally expected to outperform existing learners, even when the outcome is unconstrained, as we confirm empirically in simulation studies and an analysis of critical care data. Our development also sheds broader light onto the construction of orthogonal learners for other estimands. 

%  Please place your key words in alphabetical order, separated
%  by semicolons, with the first letter of the first word capitalized,
%  and a period at the end of the list.
%

Key-words: Causal prediction; DR-learner; Heterogeneous treatment effect; Meta-learner; Orthogonal learner; R-learner.

%  As usual, the \maketitle command creates the title and author/affiliations
%  display 

\maketitle

%  If you are using the referee option, a new page, numbered page 1, will
%  start after the summary and keywords.  The page numbers thus count the
%  number of pages of your manuscript in the preferred submission style.
%  Remember, ``Normally, regular papers exceeding 25 pages and Reader Reaction 
%  papers exceeding 12 pages in (the preferred style) will be returned to 
%  the authors without review. The page limit includes acknowledgements, 
%  references, and appendices, but not tables and figures. The page count does 
%  not include the title page and abstract. A maximum of six (6) tables or 
%  figures combined is often required.''

%  You may now place the substance of your manuscript here.  Please use
%  the \section, \subsection, etc commands as described in the user guide.
%  Please use \label and \ref commands to cross-reference sections, equations,
%  tables, figures, etc.
%
%  Please DO NOT attempt to reformat the style of equation numbering!
%  For that matter, please do not attempt to redefine anything!

\section{Introduction}
\label{s:intro}

Data-adaptive modeling (e.g., based on model selection or machine learning algorithms) is routinely used by statisticians and data scientists to quantify associations and evaluate the effects of exposures, treatments or interventions. 
A well studied example concerns estimation of the mean of a counterfactual outcome $Y^1$, which represents the outcome that would be seen for a random subject if it were treated. Under standard causal assumptions, primarily that conditioning on a measured, possibly high-dimensional collection of variables $L$ suffices to adjust for confounding of the effect of treatment $A$, coded 0 or 1,  on outcome $Y$, this can be identified as \citep{miguel2023causal}
\[E(Y^1)=E\left\{E(Y|A=1,L)\right\}.\]
Estimation may then proceed by first estimating the infinite-dimensional nuisance parameter $E(Y|A=1,L)$ using data-adaptive prediction algorithms, trained in the subsample of treated individuals, and next averaging these predictions over the entire sample (using a simple sample average).  
Such prediction algorithms - and more generally, nearly all model selection or machine learning algorithms - optimally balance bias versus variance in order to minimize expected (in-sample) prediction error. However, in doing so, they deliver no guarantees in terms of optimizing relevant performance measures (e.g., mean squared error) for the estimand of interest (e.g., for $E(Y^1)$). In fact, na\"{\i}ve use of data-adaptive strategies is well known to result in bias and excess variability \citep{newey_twicing_2004,van_der_laan_targeted_2006,robins2008higher,van_der_laan_targeted_2011,chernozhukov_double/debiased_2018}. This bias may be the result of eliminating variables that are strongly associated with the exposure of interest, thereby inducing confounding bias \citep{dukes2020obtain}; more generally, it is the result of oversmoothing in the wrong parts of the data (e.g., at the wrong covariate levels).
 
In recent years, enormous progress has been made in terms of making estimators of scalar estimands, like $E(Y^1)$, less susceptible to the bias that affects the (possibly non-parametric) data-adaptive estimators (e.g., of $E(Y|A=1,L)$) on which they are based. These developments almost exclusively rely on the so-called efficient influence curve or canonical gradient \citep{fisher2021visually,hines2021demystifying} of the considered scalar estimands. They use it either to directly de-bias na\"{\i}ve estimators \citep{chernozhukov_double/debiased_2018}, or instead to base na\"{\i}ve estimators on data-adaptive estimators that are better targeted towards the parameter of interest \citep{van_der_laan_targeted_2011}. The resulting theory is generally well developed, but limited to so-called pathwise differentiable parameters that are estimable at parametric (i.e. root-$n$) rates. It therefore does not readily extend to infinite-dimensional parameters, such as the conditional mean $E(Y^1|Z)$ of a counterfactual outcome, with $Z\subseteq L$. The estimation of such quantities is nonetheless of increasing interest for personalized decision-making assisted by  counterfactual outcome predictions \citep{athey2017beyond, kosorok2019precision}.

\cite{foster2019orthogonal} made progress in this infinite-dimensional setting by using so-called orthogonal learners. These are learners obtained by minimizing a so-called Neyman-orthogonal loss function. They do this by extending the key property of Neyman-orthogonality of influence curves to loss functions on which data-adaptive algorithms rely, with the aim to de-bias. Here, Neyman-orthogonality of a functional refers to the mean zero property of its directional derivatives \citep{foster2019orthogonal} along one-dimensional paths that (only) change one of the nuisance parameters (e.g., $E(Y|A=1,L)$ or $P(A=1|L)$) on which it is based, no matter which. A loss function is called Neyman-orthogonal when this property holds for all its directional derivatives  along one-dimensional paths that change the (infinite-dimensional) parameter of interest. The theory of \cite{foster2019orthogonal} is generic and, like \cite{chernozhukov2018plug}, provides suggestions for how Neyman-orthogonal loss functions can be constructed. However, the resulting loss functions can be difficult to optimize using off-the-shelf machine learning algorithms, a problem that we will address in this paper.

Specific orthogonal learners for $E(Y^1-Y^0|Z)$ are given in \cite{kennedy2020towards}, \cite{curth2020estimating} and \cite{nie2021quasi}, with the first two learners readily extending to the estimation of $E(Y^1|Z)$ (see later), and an additional such learner being proposed in \cite{foster2019orthogonal}. 
These learners share a number of limitations, which we aim to address here. First, it is not readily clear what these learners deliver when, because of smoothing, the loss function is optimized over a function class that does not contain the truth (see  \cite{morzywolek2023general} for further insight into this). Second, these learners have a key limitation in that they apply standard learners to transformed outcome data, which leads them to disregard the possibly constrained outcome  space. For instance, a DR-learner for $E(Y^1|Z)$ \citep{kennedy2020towards} would amount to regressing pseudo-outcomes
\[\frac{A}{P(A=1|L)}\left\{Y-E(Y|A=1,L)\right\}+E(Y|A=1,L)\]
onto $Z$, with nuisance parameters $P(A=1|L)$ and $E(Y|A=1,L)$ substituted by data-adaptive estimates.
This is particularly problematic for dichotomous outcomes since these pseudo-outcomes are not constrained to the unit interval, making it difficult for standard learners to guarantee predictions within the unit interval.

In this article we will remedy the first problem by finding the function
$m(Z)$ in some functional class $\Gamma$, e.g., the set of all (measurable) functions of $Z$ with finite second moment,
which minimises a de-biased estimator of the counterfactual prediction error
\begin{equation}\label{mse}E\left[\left\{Y^1-m(Z)\right\}^2\right].\end{equation}
A key challenge is that this de-biased estimator does not readily lend itself to minimization using off-the-shelf software for statistical/machine learning. Inspired by targeted learning algorithms \citep{van_der_laan_targeted_2011}, we therefore next target the infinite-dimensional nuisance parameters in such a way that this de-biased estimator of the loss reduces to a standard mean squared error loss w.r.t. imputed counterfactuals $Y^1$. In doing so, we address the second problem mentioned above. We show that the resulting learner is Neyman-orthogonal, and refer to it as i-learner or imputation-learner, in view of its reliance on imputed outcomes that are `orthogonalized' or targeted towards the estimation of counterfactual (conditional) means. Simulation studies and an analysis of critical care data show adequate performance, even for unconstrained outcomes.

\section{Proposal}

Consider a study design which collects i.i.d. data on a possibly high-dimensional vector of covariates $L$, that suffices to adjust for confounding of the effect of a dichotomous treatment $A$ on an outcome $Y$, in the sense that $Y^1\cip A|L$. Suppose furthermore that the consistency assumption holds that $Y^1$ equals $Y$ in distribution for individuals with $A=1$, and that $P(A=1|L)>\sigma>0$ w.p.1. Our aim is to find the function $m(Z)$ in $\Gamma$ which minimises the counterfactual prediction error (\ref{mse}).
If $\Gamma$ contains $E(Y^1|Z)$, then the above minimization problem leads to $m(Z)=E(Y^1|Z)$, but otherwise delivers the closest approximation (in mean squared error). When $Z=L$, $E(Y^1|Z)$ reduces to $E(Y|A=1,L)$ under the stated identification assumption. Often, however, we may choose $Z$ to be a small subset of $L$, either because it is logistically better feasible in clinical practice to predict $Y^1$ based on a small collection of variables, or because the uncertainty in the resulting predictions can be more accurately expressed when $Z$ is low-dimensional. In that case, the minimizer to the counterfactual prediction error can be identified as $E\left\{E(Y|A=1,L)|Z\right\}$ under the stated assumptions. 

\subsection{Minimizing counterfactual prediction error}\label{subsec:error}

The counterfactual prediction error (\ref{mse}) does not readily provide a feasible loss function for practical use. This is partly because of it being expressed in terms of population expectations, but more importantly because $Y^1$ is only measured for subjects with $A=1$. This can be remedied by instead optimizing a consistent estimator of the counterfactual prediction error. 
For this, we may use inverse probability weighting upon noting that
\begin{eqnarray*}
E\left[\left\{Y^1-m(Z)\right\}^2\right]&=&E\left[\frac{A}{g(L)}\left\{Y-m(Z)\right\}^2\right],
\end{eqnarray*}
where $g(L)=P(A=1|L)$. Alternatively, we can rewrite
\begin{eqnarray*}
E\left[\left\{Y^1-m(Z)\right\}^2\right]&=&E\left[A\left\{Y^1-m(Z)\right\}^2+(1-A)\left\{Y^1-m(Z)\right\}^2\right]\\
&=&E\left(A\left\{Y-m(Z)\right\}^2+(1-A)E\left[\left\{Y-m(Z)\right\}^2|A=1,L\right]\right).
\end{eqnarray*}
This delivers a loss function that is not easy to optimize, in view of which we rewrite 
\[E\left[\left\{Y-m(Z)\right\}^2|A=1,L\right]=\mbox{\rm Var}(Y|A=1,L)+\left\{Q(L)-m(Z)\right\}^2,\]
where $Q(L)=E\left(Y|A=1,L\right)$. Because the first term on the righthand side does not depend on $m(.)$, minimization of  (\ref{mse}) is then equivalent to minimization of 
\[E\left(A\left\{Y-m(Z)\right\}^2+(1-A)\left\{Q(L)-m(Z)\right\}^2\right).\]

The above identities suggest finding the function $m(.)$ that minimizes 
\begin{equation}\label{true}
\frac{1}{n}\sum_{i=1}^n \frac{A_i}{g(L_i)}\left\{Y_i-m(Z_i)\right\}^2+r(m),\end{equation}
or 
\begin{equation}\label{true2}
\frac{1}{n}\sum_{i=1}^n \left\{A_iY_i+(1-A_i)Q(L_i)-m(Z_i)\right\}^2+r(m),\end{equation}
where $r(.)$ is a penalty term that depends on the complexity of $m(.)$.
If $g(.)$ and $Q(.)$ were known, then this could be done by either weighting existing machine learning algorithms, or applying them to outcomes imputed as $A_iY_i+(1-A_i)Q(L_i)$. When $g(.)$ and $Q(.)$ are unknown, we may substitute them by predictions $\hat{g}(.)$ and $\hat{Q}(.)$, respectively, e.g., obtained via machine learning. We may then instead consider minimisation of
\begin{equation}\label{estimated}
\frac{1}{n}\sum_{i=1}^n \frac{A_i}{\hat{g}(L_i)}\left\{Y_i-m(Z_i)\right\}^2+r(m)
\end{equation}
or 
\begin{equation}\label{estimated2}
\frac{1}{n}\sum_{i=1}^n \left\{A_iY_i+(1-A_i)\hat{Q}(L_i)-m(Z_i)\right\}^2+r(m).\end{equation}

Unfortunately, minimisation of (\ref{estimated}) and (\ref{estimated2}) will not generally deliver an estimator of $m(.)$ that is equivalent (in large samples) to the solution to (\ref{true}) and (\ref{true2}), respectively.
For instance, for (\ref{estimated}), this is because 
\begin{eqnarray*}\frac{1}{n}\sum_{i=1}^n \frac{A_i}{\hat{g}(L_i)}\left\{Y_i-m(Z_i)\right\}^2&=&\frac{1}{n}\sum_{i=1}^n \frac{A_i}{g(L_i)}\left\{Y_i-m(Z_i)\right\}^2\\
&&+\frac{1}{n}\sum_{i=1}^n \left\{\frac{1}{\hat{g}(L_i)}-\frac{1}{g(L_i)}\right\}A_i\left\{Y_i-m(Z_i)\right\}^2,\end{eqnarray*}
where the second term in the righthand side converges to zero, but may be sizeable in finite samples when the machine learning predictions $\hat{g}(L_i)$ are slowly converging. That this may be problematic can be seen, for instance, upon choosing $m(Z_i)=Z_i'\beta$. To understand the behavior of the resulting minimizer for $\beta$, we note that it is driven by the behavior of the derivative of the above identity w.r.t. $\beta$:
\begin{eqnarray*}\frac{-2}{n}\sum_{i=1}^n \frac{A_i}{\hat{g}(L_i)}\left\{Y_i-m(Z_i)\right\}Z_i&=&\frac{-2}{n}\sum_{i=1}^n \frac{A_i}{g(L_i)}\left\{Y_i-m(Z_i)\right\}Z_i\\
&&-\frac{2}{n}\sum_{i=1}^n \left\{\frac{1}{\hat{g}(L_i)}-\frac{1}{g(L_i)}\right\}A_i\left\{Y_i-m(Z_i)\right\}Z_i.\end{eqnarray*}
The first term on the right equals $O_p(n^{-1/2})$ at the true value $\beta_0$, whereas the absolute value of each (say, the $j$th) component of the second term is upper bounded by
\[2\left[\frac{1}{n}\sum_{i=1}^n \left\{\frac{\hat{g}(L_i)-{g}(L_i)}{\hat{g}(L_i)g(L_i)}\right\}^2\right]^{1/2}
\left[\frac{1}{n}\sum_{i=1}^n A_i\left\{Y_i-m(Z_i)\right\}^2Z_{ij}^2\right]^{1/2},\]
by the Cauchy-Schwarz inequality. Here, the second term is $O_p(1)$ and the first term will generally be $O_p(n^{-b})$ for $b<1/2$ when flexible, data-adaptive methods are used for $g(.)$ (see e.g., \cite{chernozhukov_double/debiased_2018}). This can make the second term in the above expansion dominant, causing the minimizer of (\ref{estimated}) to be further than the typical $n^{1/2}$ distance away from $\beta_0$, despite the use of a parametric model for $m(.)$. Likewise, when more general data-adaptive meta-learners are used for $m(.)$, their convergence rate may be harmfully affected by slow convergence in $\hat{g}(.)$ \citep{foster2019orthogonal,kennedy2020towards,nie2021quasi}, which may well be much slower than the convergence rate of $\hat{m}(.)$ (at known nuisance parameters) when $L$ is of higher dimension than $Z$.

\subsection{Constructing orthogonal loss functions}

The above concerns can be remedied by instead minimizing a double robust estimator of the (empirical) counterfactual prediction error \citep{foster2019orthogonal,coston2020counterfactual}:
\[\frac{1}{n}\sum_{i=1}^n \frac{A_i}{\hat{g}(L_i)}\left(\left\{Y_i-m(Z_i)\right\}^2-\hat{E}\left[\left\{Y_i-m(Z_i)\right\}^2|A_i=1,L_i\right]\right)+\hat{E}\left[\left\{Y_i-m(Z_i)\right\}^2|A_i=1,L_i\right].\]
By the earlier remarks, this is equivalent to minimization of 
\begin{eqnarray}
&&\frac{1}{n}\sum_{i=1}^n \frac{A_i}{\hat{g}(L_i)}\left[\left\{Y_i-m(Z_i)\right\}^2-\left\{\hat{Q}(L_i)-m(Z_i)\right\}^2\right]+\left\{\hat{Q}(L_i)-m(Z_i)\right\}^2\nonumber\\
&&=\frac{1}{n}\sum_{i=1}^n \frac{A_i}{\hat{g}(L_i)}\left\{Y_i-m(Z_i)\right\}^2+\left\{1-\frac{A_i}{\hat{g}(L_i)}\right\}\left\{\hat{Q}(L_i)-m(Z_i)\right\}^2\label{dr}\\
&&=\frac{1}{n}\sum_{i=1}^n\left\{A_iY_i+(1-A_i)\hat{Q}(L_i)-m(Z_i)\right\}^2\nonumber\\&&+A_i\frac{1-\hat{g}(L_i)}{\hat{g}(L_i)}\left[\left\{Y_i-m(Z_i)\right\}^2-\left\{\hat{Q}(L_i)-m(Z_i)\right\}^2\right]\nonumber,
\end{eqnarray}
where minimization of the latter is equivalent to minimization of 
\begin{eqnarray}
&&\frac{1}{n}\sum_{i=1}^n\left\{A_iY_i+(1-A_i)\hat{Q}(L_i)-m(Z_i)\right\}^2-A_i\frac{1-\hat{g}(L_i)}{\hat{g}(L_i)}\left\{Y_i-\hat{Q}(L_i)\right\}2m(Z_i)\label{dr2}.
\end{eqnarray}
Here, (\ref{dr}) shows that this double robust estimator updates the inverse probability weighted loss (\ref{estimated}) to include also data for the unexposed (i.e., those with $A_i=0$), thereby increasing efficiency and robustness to a possible lack of consistency of $\hat{g}(.)$ (provided that a consistent and sufficiently fast converging estimator $\hat{Q}(.)$ is used). Likewise, (\ref{dr2}) shows that this double robust estimator updates the regression imputed loss (\ref{estimated2}) to increase robustness to a possible lack of consistency of $\hat{Q}(.)$ (provided that a consistent and sufficiently fast converging estimator $\hat{g}(.)$ is used).
That minimization of (\ref{dr}) makes the meta-learner less sensitive to the estimation of nuisance parameters follows from \cite{foster2019orthogonal} (see also \cite{morzywolek2023general}); the crux of the idea is summarized in Web Appendix A. In particular, double robustness of the loss makes it Neyman-orthogonal - referred to as `orthogonal' hereafter - in the sense that its directional derivatives w.r.t. $m(.)$ and one of the nuisance parameters (e.g., $E\left(Y_i|A_i=1,L_i\right)$ or $P(A_i=1|L_i)$) have mean zero (at the truth).

\subsection{Minimizing the orthogonal loss}

Unfortunately, standard learners do not readily lend themselves towards minimization of (\ref{dr}) and (\ref{dr2}).
\cite{kennedy2020towards} remedies this by instead regressing the pseudo outcome
\[\frac{A_i}{\hat{g}(L_i)}\left\{Y_i-\hat{Q}(L_i)\right\}+\hat{Q}(L_i)\]
onto $Z_i$ using a standard learner, i.e., by minimizing 
\[\frac{1}{n}\sum_{i=1}^n\left[\frac{A_i}{\hat{g}(L_i)}\left\{Y_i-\hat{Q}(L_i)\right\}+\hat{Q}(L_i)-m(Z_i)\right]^2+r(m).\]
Related, but different proposals are given in \cite{nie2021quasi}, \cite{curth2020estimating} and \cite{morzywolek2023general}. A limitation of some proposals is that it is not always readily clear what counterfactual loss (e.g., (\ref{mse})) they are aiming to minimize (see \cite{morzywolek2023general} for further discussion on this point). Moreover, 
unlike (\ref{dr}), the above loss function contrasts the predictions $m(Z_i)$ with transformed outcomes, which may not belong to the same outcome space. This is especially problematic for dichotomous exposures, as it makes it difficult to guarantee estimated outcome probabilities in the unit interval. This is likewise the case for the above alternative proposals.

When $Z_i=L_i$, this problem is readily accommodated by minimizing (\ref{estimated})
instead.
The reason is that, interestingly, this loss is orthogonal when $Z_i=L_i$, though not otherwise. This can be seen because its directional derivative w.r.t. $m(Z_i)$ and then $\hat{g}(L_i)$ equals 
\[\frac{1}{n}\sum_{i=1}^n A_i\left\{Y_i-m(Z_i)\right\}\theta(L_i)\]
for some function $\theta(L_i)$; this has mean zero at the truth, since then $m(Z_i)=E(Y_i|A_i=1,L_i)$. 
Informally, the reason that this works is that $m(.)$ equals $Q(.)$ in that case, so that the double robustness guarantee can also be achieved via $m(.)$ rather than $Q(.)$.  
Note that this is not the case when instead minimizing (\ref{estimated2}), which can be informally seen because it ignores the propensity score,
 or when minimizing
\[\frac{1}{n}\sum_{i=1}^n A_i\left\{Y_i-m(Z_i)\right\}^2+r(m),\]
which instead targets minimization of $E\left[\left\{Y^1-m(Z)\right\}^2|A=1\right]$.
In the next section, we will generalize this ad hoc solution (i.e., minimization of  (\ref{estimated})) to make it work also when $Z_i\subseteq L_i$. We will achieve this by targeting the estimation of the nuisance parameters.

\subsection{Targeting the nuisance parameter learners}

We will find a solution to the above problem by minimizing (\ref{estimated2}) in a way that makes it equivalent to minimizing (\ref{dr2}). For this, we wish the term
\begin{equation}\label{targeting}
\frac{1}{n}\sum_{i=1}^n A_i\frac{1-\hat{g}(L_i)}{\hat{g}(L_i)}\left\{Y_i-\hat{Q}(L_i)\right\}m(Z_i)\end{equation}
in (\ref{dr2}) to be sufficiently close to zero for `all' functions $m(.)$ in some function class (considering that $m(.)$ is unknown). While it is generally close to zero (as a result of averaging contributions $Y_i-\hat{Q}(L_i)$ with mean zero (conditional on $L_i$) in large samples), it is not generally close enough. In view of this, we will target or update the obtained predictions $\hat{Q}(L_i)$ to shrink (\ref{targeting}) closer to zero, so that minimization of (\ref{estimated2}) based on the targeted predictions $\hat{Q}(.)$ is asymptotically equivalent to minimization of the orthogonal loss (\ref{dr2}). This is inspired by targeted learning algorithms \citep{van_der_laan_targeted_2011}, but here necessitates targeting in infinitely many directions over the function class of $m(.)$; targeting in infinitely many directions was previously considered in a longitudinal context in 
\cite{luedtke2017sequential}. 
We will refer to the resulting learner, which minimizes (\ref{estimated2}) based on the targeted predictions, as i-learner or imputation-learner.

Targeting $\hat{Q}(L_i)$ (i.e., updating initial estimates of $\hat{Q}(L_i)$) so that (\ref{targeting}) is close to zero for all functions $m(.)$ is challenging by the fact that $m(.)$ is unknown. We will therefore make 2 assumptions. First, we will make a sparsity assumption that $m(.)$ depends only on $D\leq d$ components of $Z$. Second, we will impose a smoothness assumption that $m(.)$ obeys a Tensor product space model \citep{lin2000tensor}, which postulates that $m(.)$ can be written as a finite sum of $d$-dimensional products (with $d$ the dimension of $Z$) of univariate functions in a first-order Sobolev space (i.e., functions that are absolutely continuous and have a first order derivative that is mean square integrable). This is approximately equivalent to assuming that all mixed first order derivatives of $m(.)$ are mean square integrable \citep{zhang2022regression}. This smoothness assumption enables us to write $m(.)$ as an
infinite linear combination of a dictionary of (orthonormal) basis functions $b(.)$, i.e., 
\[m(Z)=\sum_{j=1}^{\infty} b_j(Z)\gamma_j,\]
with coefficients $\gamma_j$ that decay at faster rate than $j^{-1.5}$ so that 
\[\sum_{j=1}^{\infty} \left(\frac{j}{\max{(\log^{D-1}{j},1)}}\right)^{2}\gamma_j^2\leq Q\]
for some constant $Q$. For this, we will use products of finite numbers of univariate cosine basis functions (including the constant function 1) to construct a richer function class (with unravelling rules as detailed in \cite{zhang2022regression} to (a) impose an ordering on the basis functions whereby lower order terms are prioritized and (b) exclude products of functions of more than $D'$ variables, where $D'$ is chosen by the user and assumed to exceed $D$). Note that when these restrictions fail to hold, then by the choice of loss function, the i-learner will still provide the function with the smallest counterfactual prediction error in the considered function class.

This exponential decay of the coefficient series $\gamma_j$ justifies 
approximating $m(Z)$ with a truncated series 
\[\sum_{j=1}^{J_n} b_j(Z)\gamma_j,\]
so that the problem of shrinking (the absolute value of) (\ref{targeting}) with $m(.)$ unknown simplifies to that of shrinking (the absolute value of)
\[\frac{1}{n}\sum_{i=1}^n A_i\frac{1-\hat{g}(L_i)}{\hat{g}(L_i)}\left\{Y_i-\hat{Q}(L_i)\right\}b_j(Z_i)\]
for $j=1,...,J_n$ with $b_j(.)$ known. For this, 
 we will build a parametric submodel around initial predictions $\hat{Q}^{(0)}(L_i)$
as follows:
\[h\left\{E(Y_i|A_i=1,L_i)\right\}=h\left\{\hat{Q}^{(0)}(L_i)\right\}+\epsilon'b(Z_i)\frac{1-\hat{g}(L_i)}{\hat{g}(L_i)}\]
for a link function $h(.)$ that is the identity function $(h(x)=x$) for a continuous outcome, or the logistic function ($h(x)=\mbox{\rm logit}(x)$)
for a dichotomous outcome, and fit the model using maximum likelihood
with $l_1$-penalisation with lasso penalty of the standard order $\sqrt{\log{(J_n)}/n}$ with $J_n=C(D)d^{D'}n^{1/3}\log^{D'-1}{(n)}$ with $C(D)$ a constant that may depend on $D$.
In Web Appendix B, we show for $h(.)$ the identity function that this procedure indeed achieves the required shrinkage. In particular, we confirm that the term (\ref{targeting}) shrinks from being of the order $O_p(n^{-b})$ to being:
\begin{equation}\label{shrink}
O_p\left(\log^{\frac{D-1}{3}}{n} \left\{\log{J_n}n^{-2/3}\log^{\frac{D-1}{3}}{n}
+ n^{-5/6}+n^{-b-1/3}\right\}\right),\end{equation}
Here, the second 2 terms reflect approximation error as a result of approximating $m(.)$ by means of a finite number, $J_n$, of basis functions. It shows that the earlier root mean squared prediction error in $\hat{Q}^{(0)}$ of the order $O_p(n^{-b})$ shrinks by a factor $n^{-1/3})$, up to polylog terms (note also that the effective dimension $D$ only shows up on the exponent of the log sample size, rather than the sample size itself). The first term arises from not knowing which $D$ out of $d$ terms to select and is of the order $O_p(n^{-2/3})$, up to polylog terms. This term is not dependent on the convergence rate of $\hat{Q}^{(0)}$, as the procedure is effectively trying to set (\ref{targeting}) to zero no matter the choice of $\hat{Q}^{(0)}$, and that it depends only on the logarithm of the ambient dimension $d$.

In Web Appendix B, we further study the mean squared error $E\left[\left\{\hat{m}(Z)-m(Z)\right\}^2\right]$ of the resulting estimator $\hat{m}(Z)$. We find it to be generally dominated by 
the oracle excess risk (if nuisance parameters were known),
which we expect to be of the order $O_p(n^{-2/3})$, up to polylog terms (see lemma C.6 in \cite{zhang2022regression}). 
% note that I refer to this lemma instead of Theorem 5.1 because this shows the excess risk and not the generalization error
In particular, the mean squared error is upper bounded by the excess risk plus 
(\ref{shrink}), which is generally of the same order of magnitude, and plus a product rate term involving the mean squared errors of the nuisance parameter estimators $\hat{g}(.)$ and $\hat{Q}(.)$.
This confirms that optimization of (\ref{dr2}) roughly delivers oracle behavior, and thus that the considered targeting step suffices. In particular, the slow convergence rates in the nuisance parameter estimators are attenuated (by the product rate term, which results from the loss function being approximately Neyman-orthogonal, and by the targeting step) before propagating into the meta-learner.
One exception is when parametric regression is used for $m(.)$; however, in that case, one can make (\ref{targeting}) exactly zero by letting $b(Z)$ be the derivatives of $m(Z)$ w.r.t. the parameters indexing its parametric model (without the need for $l_1$-penalization).

\subsection{Cross-fitting} \label{subsec:crossfit}

As in \cite{foster2019orthogonal}, \cite{kennedy2020towards} and \cite{nie2021quasi}, sample-splitting is crucial for the i-learner to perform well. It is needed to be able to invoke the generalization error bounds for the machine learning algorithm used for the plug-in empirical risk minimization, since it requires the data to be independent and identically distributed (i.i.d.). 
% note that in that case we look at sample averages MSE instead of population MSE
Fitting the nuisance parameters on the same sample as used for minimizing the orthogonal loss would result in minimizing the loss over a correlated sample (through the plug-in estimates of the nuisance parameters) and hence would lead to violation of the i.i.d. assumption. To prevent efficiency loss, we performed sample splitting via a cross-fitting procedure as follows. First, we split the data on the $K$ (e.g. 5) disjoint, roughly equally sized folds. Then for each fold $k=1,...,K$, we train $E(A|L)$ and $E(Y|A=1,L)$ on the data from the $K-1$ remaining folds, and subsequently target the predictions for $E(Y|A=1,L)$ on those same folds (we chose not to do the targeting step on the $k$th fold in view of the high-dimensionality of the models used for targeting). We then used the resulting predictions for $E(Y|A=1,L)$ to calculate imputed outcomes in the $k$th fold. After having repeated this for all folds, we regressed the resulting imputed outcomes on the covariates $Z$ across all folds in one go.

%LIBRARY FOREACH, DOPARALLEL

\section{Simulation study}

We have evaluated the performance of the proposal in two simulation experiments.
Our first simulation study focuses on continuous outcomes using the following data generating mechanism (largely) from \cite{kang2007demystifying} to generate i.i.d. data: $L_i \sim \mathcal{N} \left( 0,  \Sigma_{d \times d} \right)$, where $\Sigma_{d \times d}$ is a $d$-dimensional correlation matrix with $d=20$ and correlations drawn from a normal distribution and varying from -0.5 to 0.5, $A_i \vert L_i \sim \text{Bern} \left\lbrace \pi \left( L_i \right) \right\rbrace$ with $\pi \left( L_i \right) = 1 / \left[1+\exp\left\{V_1 - 0.5V_2 + 0.25V_3 + 0.1V_4\right\}\right]$ with $V_1=\exp(L_1/2)$, $V_2=L_2/\left\{1+\exp(L_1)\right\}+10$, $V_3=(L_1L_3/25+0.6)^3$ and $V_4=(L_2+L_4+20)^2$, $Y_i=Y^a_i \vert L_i \sim N\left\lbrace  b \left( L_i \right),1 \right\rbrace$ for $a=0,1$, where $b \left( L_i \right) = 210+27.4V_1+13.7V_2+13.7V_3+13.7V_4$. 
Our second simulation study focuses on dichotomous outcomes, using the following data generating mechanism (largely) from \cite{webdgm}. Covariates $L_i$ were generated as before, $A_i \vert L_i \sim \text{Bern} \left\lbrace \pi \left( L_i \right) \right\rbrace$ with $\pi \left( L_i \right) = 1 / \left[1+\exp\left\{2+\sin(K_i)+\cos(K_i)\right\}\right]$ with $K_i=\sum_{j=1}^p L_{ij}/j$ and $Y_i=Y^a_i \vert L_i \sim \text{Bern} \left\lbrace  b \left( L_i \right) \right\rbrace$ for $a=0,1$, where $b \left( L_i \right) = 1/\left[1+\exp\left\{-2.5+2\cos^2(K_i)\right\}\right]$.

In each of 500 replications, for different estimators $\hat{m}(.)$ of ${m}(.)$, the simulation studies evaluate the mean squared errors 
\[\frac{1}{500}\sum_{i=1}^{500}\left\{\hat{m}(Z_i^{(v)})-{m}(Z_i^{(v)})\right\}^2\]
based on a random validation sample of observations $Z_i^{(v)}$, $i=1,...,500$, drawn from the same data-generation model as specified above. We then averaged mean squared errors over the 500 simulation experiments. Specifically, we considered the following learners of $E(Y^1|Z)$, all based on $l_1$-penalized sieves with cosine basis, 5-fold cross-fitting and `standard' learners (to be specified later) for the nuisance parameters: (1) na\"{\i}ve: sieves fitted in the treated subgroup; (2) IPW: sieves fitted in the treated subgroup, using inverse probability weighting as in (\ref{estimated}); (3) imputation: sieves based on regression mean imputation, as in (\ref{estimated2}); (4) DR: DR-learner; (5) i-learner: sieves based on targeted regression mean imputations. As standard learners for the estimation of nuisance parameters, we considered regression forests as well as SuperLearner with a library given by generalized linear main effect and interaction models, stepwise regression, regression forests and generalized additive models. We repeated this for $Z_i$ equalling the first 2 covariates in $L_i$, the first 5, and finally all 20. Note that DR-learner is based on unconstrained pseudo-outcomes and that estimator 1 is expected to be biased as a result of ignoring confounding. Estimators 2 and 3 are based on non-orthogonal learners (except estimator 2 when $V_i=L_i$) and may therefore also be subject to bias. All other learners are orthogonal. 

\begin{table}[htbp]
\caption{Results of simulation experiment 1 (continuous outcome). Choice of learner for the nuisance parameters, the covariate dimension for the meta-learner, the number of basis functions used in the penalized sieve estimator, mean squared errors of the 5 listed meta-learners.}
\centering
\begin{tabular}{cccccccc}\hline
Learner & dim($Z$) & nr basis fct. & na\"{\i}ve & IPW & imputation & DR & i-learner \\\hline
Random forest & 2 & 10 & 495 & 495 & 480 & 471 & 466\\
& 5 & 20 & 105 & 100 & 128 & 102 & 120\\ 
& 20 & 50 & 46 & 47 & 117 & 87 & 110 \\
SuperLearner & 2 & 10 & 495 & 497 & 446 & 448 & 446\\
& 5 & 20 & 105 & 102 & 39 & 43 & 39 \\
& 20 & 50 & 47 & 49 & 13 & 21 & 13 \\\hline
\end{tabular}
\end{table}

The results from the first simulation experiment with nuisance parameters estimated via random forests show favourable performance of the orthogonal meta-learners (DR and i-learner) when $Z$ is low-dimensional, with the DR-learner being best, but worse behaviour at higher-dimensional $Z$. These results may not perfectly align with the theory, which assumes `reasonable' convergence of the random forest regression fits; this is unlikely met in this complex data-generating mechanism. We therefore see a different picture when the SuperLearner is used for nuisance parameter estimation, with the orthogonal learners drastically outperforming the na\"{\i}ve and IPW-learner, and the proposed estimator being uniformly best. The second simulation experiment uses nuisance parameters that were previously observed to be well estimable using random forest regression \citep{webdgm}. We therefore see quite similar behaviour as with SuperLearner, except that the additional flexibility of SuperLearner leads to instability in the DR-learner and more estimates outside the unit interval. The proposed estimator becomes relatively favourable, but is outperformed by the imputation estimator without targeting. While the performance of this estimator cannot be guaranteed based on theoretical results (as is known for related g-computation estimators based on data-adaptive nuisance parameter estimates), one cannot exclude that it may perform better in some data-generating mechanisms.   

\begin{table}[htbp]
\caption{Results of simulation experiment 2 (dichotomous outcome). Choice of learner for the nuisance parameters, the covariate dimension for the meta-learner, the number of basis functions used in the penalized sieve estimator, mean squared errors of the 5 listed meta-learners, and the percentage of DR-learner estimates outside of  the unit interval ($\%$).}
\centering
\begin{tabular}{cccccccccc}\hline
Learner & dim($Z$) & nr basis fct. & na\"{\i}ve & IPW & imputation & DR & i-learner & \%\\\hline
Random forest & 2 & 10 & 0.020 & 0.020 & 0.012 & 0.018 & 0.017 & 2.9\\
& 5 & 20 & 0.021 & 0.020 & 0.011& 0.019& 0.016 & 4.1 \\ 
& 20 & 50 & 0.022 & 0.021 & 0.012 & 0.019 & 0.016 & 2.2\\
SuperLearner & 2 & 10 & 0.020 & 0.024 & 0.013 & 0.034 & 0.018 & 6.9\\
& 5 & 20 & 0.020 & 0.026 & 0.012& 0.037& 0.016 & 6.7 \\ 
& 20 & 50 & 0.022 & 0.026 & 0.014 & 0.032 & 0.017 & 5.0\\\hline
%Random forest & 2 & 10 & 0.025 & 0.026 & 0.014 & 0.024 & 0.021 & 5.0\\
%& 5 & 20 & 0.026 & 0.026 & 0.014& 0.025& 0.019 & 4.5 \\ 
%& 20 & 50 & 0.028 & 0.027 & 0.014 & 0.024 & 0.019 & 2.6
%n & dim($Z$) & 1 (na\"{\i}ve) & 2 (IPW) & 3 (DR) & 4 (imp) & 5 (AIC) & 6 (lasso) & 7 (sieve) & \%\\\hline
%500 & 2 & 0.190 & 0.190 & 0.261 & 0.178 & 0.175 & 0.187 & 0.179 & 25.4 \\
%& 5 & 0.189 & 0.189 & 0.219 & 0.176 & 0.174& 0.188 & 0.179 & 16.5\\
%& 20 & 0.188 & 0.188 & 0.203 & 0.178 & 0.179 & 0.189 & 0.178 & 8.2\\
%1000 & 2 & 0.189 & 0.189 & 0.250 & 0.178 & 0.175 & 0.185 & 0.181 & 22.9\\
%& 5 & 0.189 & 0.188 & 0.208 & 0.176 & 0.174 & 0.186 & 0.181 & 13.3\\
%& 20 & 0.188 & 0.187 & 0.194 & 0.179 & 0.180 & 0.186 & 0.179 & 6.4\\\hline
\end{tabular}

%%% JULY
%>     cat(j,apply(mse[1:j,c(1:5,16)],2,mean,na.rm=T),"\n")
%500 0.02477562 0.02567063 0.01354822 0.0238277 0.02069395 0.05011623 
%>     cat(j,apply(mse[1:j,c(6:10,17)],2,mean,na.rm=T),"\n")
%500 0.02567367 0.02589778 0.01360598 0.02503712 0.01906307 0.04478557 
%>     cat(j,apply(mse[1:j,c(11:15,18)],2,mean,na.rm=T),"\n")
%500 0.02820996 0.02718292 0.01397801 0.02350839 0.01928188 0.02550301 

% 500
%500 0.1623578 
%500 0.1902174 0.1869625 0.2613474 0.1777799 0.1900471 0.1791379 0.1748122 0.254 
%500 0.1891341 0.1884408 0.219463 0.1764891 0.1894308 0.1791099 0.1743329 0.165 
%500 0.1882251 0.1892096 0.2033078 0.1783875 0.1885876 0.1779247 0.1791685 0.082 

%1000
%0.1664891 
%500 0.1885605 0.1852887 0.2496706 0.1783518 0.1886253 0.1810461 0.1745085 0.229 
%500 0.1882217 0.1859645 0.2080009 0.1758711 0.1889629 0.1806886 0.1740696 0.1325 
%500 0.1870498 0.1859403 0.1939914 0.179193 0.1876917 0.1787624 0.1796703 0.064
\end{table}

% my_vector <- c(10.93116, 10.58435, 10.66843, 10.56542, 12.04772, 13.0815, 10.64859, 7.927514, 7.893729, 7.922417, 7.864557, 8.544131, 8.391265, 7.856437, 8.824958, 8.142789, 8.562223, 8.125383, 9.664171, 8.263618, 8.295405)
%[c(5, 1, 4, 3, 7, 2, 6, 12, 8, 11, 10, 14, 9, 13, 19, 15, 18, 17, 21, 16, 20)]
% my_matrix <- matrix(my_vector,nrow=3,ncol=7,byrow=T)

%\textit{These are results for simulation study 1. Quick conclusion for now is that improves upon DR-learner. Remember that in this study no improvement was necessarily expected, and we merely wanted to see no worse results than for DR-learner. The real benefits remain to be seen for dichotomous outcomes.}
%\begin{verbatim}
 %         [,1]      [,2]      [,3]      [,4]      [,5]      [,6]      [,7]
%[1,] 12.047720 10.931160 10.565420 10.668430 10.648590 10.584350 13.081500
%[2,]  8.544131  7.927514  7.864557  7.922417  7.856437  7.893729  8.391265
%[3,]  9.664171  8.824958  8.125383  8.562223  8.295405  8.142789  8.263618
%\end{verbatim}

\section{Causal prediction in critical care}

Acute kidney injury (AKI) is an abrupt decrease in kidney function, which is commonly defined in terms of KDIGO criteria \citep{KDIGO2012}. Renal replacement therapy (RRT) is a treatment that is commonly used for the management of critically ill patients with severe AKI, in particular those experiencing metabolic or fluid-related complications. RRT may rapidly correct some of the life-threatening complications associated with AKI, e.g., severe hyperkalaemia (i.e., serum potassium above $6.0$mmol/L), metabolic acidosis (i.e., pH below $7.2$) or pulmonary oedema (i.e., abnormal accumulation of fluid in lungs due to fluid overload). However, it is a very invasive treatment and may put treated patients at risk of bleeding, infection, hemodynamic instability, electrolyte abnormalities, ... It is therefore of paramount importance to carefully and appropriately judge the costs and benefits of initiating such an invasive intervention. As part of the development of a decision support system, we are therefore interested in predicting the $7$-day ICU mortality under initiation of RRT within $24$h from the time of stage $2$ AKI diagnosis in the stage $\geq2$ AKI patient population (i.e., the potential outcome $Y^1$), and the corresponding $7$-day ICU mortality under no initiation of RRT (i.e., $Y^0$).
For this, we analyzed data from the Intensive Care Information System of the Ghent University Hospital ICUs, which contains records from all adult patients admitted to the intensive care unit since $2013$. In our analysis we considered $3728$ adult stage $2$ and $3$ AKI patients admitted to the ICU between $1/1/2013$ and $31/12/2017$, who had no recorded RRT history and no RRT restrictions by the time of the inclusion at stage $2$ AKI diagnosis. 

For each patient the database holds information on several characteristic, e.g., ICU admission time, ICU discharge time, vital status at discharge, timestamps of all dialysis sessions during each ICU episode, baseline covariates (e.g., age, weight, gender, admission category \{"No surgery", "Planned surgery", "Emergency surgery"\}, receipt of dialysis prior to current ICU admission, chronic kidney disease diagnosis prior to current ICU admission) and longitudinal measurements over the  ICU episode (e.g., SOFA scores, having reached KDIGO AKI (stage $1/2/3$) creatinine condition, having reached KDIGO AKI (stage $1/2/3$) oliguric condition, receipt of diuretics, cumulative total fluid intake, cumulative total fluid output, arterial pH, serum potassium (in mmol/L), serum ureum (in mg/dL), serum magnesium (in mmol/L), fraction of inspired oxygen (FiO$_2$), peripheral oxygen saturation (SpO$_2$), arterial oxygen concentration (PaO$_2$), ratio of arterial oxygen concentration to the fraction of inspired oxygen (P/F ratio), DNR ("Do Not Resuscitate") code) and their timestamps.

We applied several meta-learners based on $l_1$-penalized sieve regression with cosine basis in order to predict potential outcomes based on the following subset of covariates $Z$: age on admission, gender, serum potassium, arterial pH, cumulative total fluid intake and cumulative total fluid output. In particular, we implemented the `na\"{\i}ve' approach, the IPW-learner,  the DR-learner, the proposed i-learner and its non-targeted equivalent. The models for the nuisance parameters have been computed using \texttt{SuperLearner} \citep{VanDerLaan2007} with the following list of wrappers: glm, glmnet, random forest (\texttt{ranger}) and xgboost.
To perform our analysis, we split the data into three equally sized parts: training set A, training set B and a test set. 
To make the most efficient use of the available data we apply cross-fitting as described in Section \ref{subsec:crossfit}. In the final step, given the models for the mean potential outcomes obtained via cross-fitting on the training data sets A and B, we use the test data set to obtain the final output, i.e., $7$-day ICU mortality under initiation of RRT (and similarly under no initiation of RRT) within 24h from stage $2$ AKI diagnosis, conditional on the selected patient characteristics. We evaluate performance on the test set to avoid possible overoptimism, which could arise once evaluating the performance of different methods on the same data that has been used for training the models.

\begin{figure}[ht!]
\begin{center}
\includegraphics[width=\textwidth]{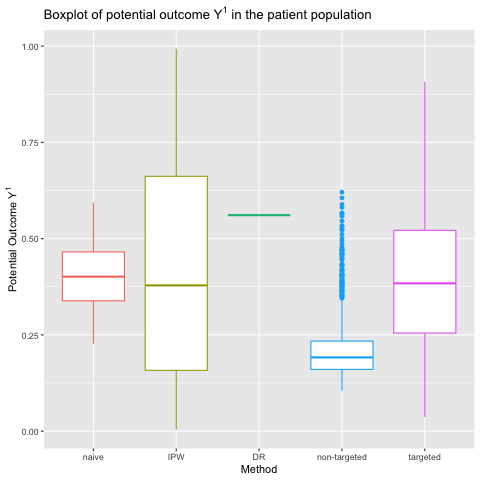}
\end{center}
\caption{Boxplot of the $7$-day ICU mortality under initiation of RRT within $24$h from the stage $2$ AKI diagnosis in the stage $\geq2$ AKI patient population, i.e., the potential outcome $Y^1$, conditional on the values of age on admission, gender, serum potassium, arterial pH, fluid intake and fluid output computed in the whole patient population using "naive" approach ("naive"), IPW-learner ("IPW"), DR-learner ("DR"), proposed targeted learner, i.e. i-Learner ("targeted"), and its non-targeted equivalent ("non-targeted").} \label{figure3}
\end{figure}

\begin{figure}[ht!]
\begin{center}
\includegraphics[width=\textwidth]{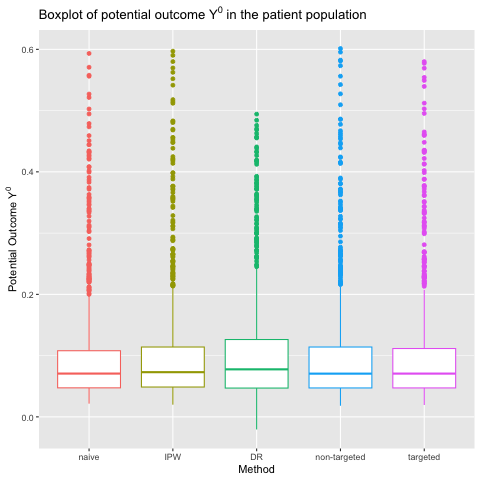}
\end{center}
\caption{Boxplot of the $7$-day ICU mortality without initiation of RRT within $24$h from the stage $2$ AKI diagnosis in the stage $\geq2$ AKI patient population, i.e., the potential outcome $Y^0$, conditional on the values of age on admission, gender, serum potassium, arterial pH, fluid intake and fluid output computed in the whole patient population using "naive" approach ("naive"), IPW-learner ("IPW"), DR-learner ("DR"), proposed targeted learner, i.e. i-Learner ("targeted"), and its non-targeted equivalent ("non-targeted").} \label{figure4}
\end{figure}

Figure \ref{figure3} presents a boxplot of the $7$-day ICU mortality estimates under initiation of RRT within $24$h from stage $2$ AKI diagnosis in stage $\geq2$ AKI patients. It shows the poor performance of DR-learner, which is the result of extreme propensity scores for some patients, making the pseudo-outcomes highly variable. In Web Appendix D, we show results for sieves with numbers of basis functions different from the default in the \texttt{Sieve} package \citep{zhang2022regression}. It shows the lack of stability of some learners (in particular, IPW-learner and DR-learner) as opposed to the proposed i-learner. Figure \ref{figure4} shows analogous results for the $7$-day ICU mortality without initiation of RRT within $24$h from the stage $2$ AKI diagnosis in the stage $\geq2$ AKI patient population. Results are more comparable between learners because the majority of patients was not treated within 24h from AKI-diagnosis. 

\section{Discussion}

We have shown how non-orthogonal learners can be made approximately Neyman-orthogonal by invoking infinite-dimensional targeting procedures, akin to iTMLE \citep{luedtke2017sequential}. 
This is justified according to our asymptotic approximations in terms of regret bounds, as well as confirmed to work well empirically. 
This result is useful because orthogonal learning procedures often demand optimizing loss functions that are difficult to optimize. Popular learners, such as DR-learner and R-learner, overcome this by applying standard learners to suitably transformed outcomes, possibly additionally invoking well-chosen observation weights. However, this comes at the expense of sub-optimal performance. We have instead remedied this by applying standard learners to carefully imputed (counterfactual) outcomes.

Our proposal provides a natural alternative for DR-learner, which likewise aims to minimize counterfactual prediction error \citep{morzywolek2023general}. It remains to be seen how it extends to an R-learner-like and a `treatment effect in the untreated' strategy for counterfactual prediction, which according to the findings in \cite{morzywolek2023general}, would naturally focus on minimizing
\[E\left[w(L)\left\{Y^1-m(Z)\right\}^2\right],\]
with $w(L)=P(A=1|L)P(A=0|L)$ or $w(L)=P(A=0|L)$, respectively.
This could likewise be done by minimizing 
\[\frac{1}{n}\sum_{i=1}^n \hat{g}(L_i)\left\{1-\hat{g}(L_i)\right\}\left\{A_iY_i+(1-A_i)\hat{Q}(L_i)-m(Z_i)\right\}^2+r(m),\]
but necessitates a different targeting step, which we will develop in future work. Such further extension will be essential for our motivating application: there are many patients for whom RRT is not a meaningful treatment strategy, which makes it suboptimal to minimize counterfactual prediction error over the full stage $\geq2$ patient population.

A further limitation of our proposal is that it imposes sparsity assumptions on the counterfactual means $E(Y^1|Z)$ and $E(Y^0|Z)$. Sparsity assumptions may more likely hold w.r.t. the conditional causal effect $E(Y^1-Y^0|Z)$ because causal effects may often be homogeneous, small or even absent; in such cases, differences between our current predictions for $Y^1$ and $Y^0$ may show erratic behaviour by not leveraging smoothness/sparsity assumptions directly on $E(Y^1-Y^0|Z)$ \citep{kunzel2019metalearners}. 
In future work, we will therefore study how this can be done when the aim is to predict both counterfactuals. We will then also develop insight into the number and choice of basis functions \citep{yang2023forster}, and study how well the proposed targeting procedure continues to work when the targeting step is based on penalized sieves, but the final optimization of the loss function is based on more general learning procedures (e.g., random forest regression). Finally, drawing inference based on the obtained infinite-dimensional estimates $E(Y^1|Z)$ and $E(Y^0|Z)$ is a challenging problem that has received little attention so far, and for which we will study the use of debiased lasso \citep{xia2023debiased}, as well as more generic procedures \citep{luedtke2023one}.

\bibliographystyle{apalike} %apalike
\bibliography{CDRLearner}

\label{lastpage}

\section*{Appendix A}

In this Appendix, we develop oracle bounds for the (orthogonal) meta-learner which minimizes the double-robust loss
\[{\cal L}(m,\eta)\equiv E\left[\frac{A}{g(L)}\left\{Y-m(Z)\right\}^2+\left\{1-\frac{A}{g(L)}\right\}\left\{Q(L)-m(Z)\right\}^2\right]\]
for $\eta\equiv (g,Q)$.

A Taylor series expansion using directional derivatives \citep{foster2019orthogonal} shows that
\begin{eqnarray*}
{\cal L}(\hat{m},\hat{\eta})&=&{\cal L}(m,\hat{\eta})
-2E\left(\left[\frac{A}{\hat{g}(L)}\left\{Y-\hat{Q}(L)\right\}+\hat{Q}(L)-m(Z)\right]\left\{\hat{m}(Z)-m(Z)\right\}\right)\\
&&+E\left[\left\{\hat{m}(Z)-m(Z)\right\}^2\right]
\end{eqnarray*}
from which 
\begin{eqnarray*}
E\left[\left\{\hat{m}(Z)-m(Z)\right\}^2\right]&=&{\cal L}(\hat{m},\hat{\eta})-{\cal L}(m,\hat{\eta})\\&&
+2E\left(\left[\frac{A}{\hat{g}(L)}\left\{Y-\hat{Q}(L)\right\}+\hat{Q}(L)-m(Z)\right]\left\{\hat{m}(Z)-m(Z)\right\}\right)
\end{eqnarray*}
Here, the expectation in last second term can be further expanded as 
\begin{eqnarray*}
&&E\left(\left[\frac{A}{{g}(L)}\left\{Y-{Q}(L)\right\}+{Q}(L)-m(Z)\right]\left\{\hat{m}(Z)-m(Z)\right\}\right)\\
&&-E\left(\left[\frac{A}{{g}^2(L)}\left\{Y-{Q}(L)\right\}\right]\left\{\hat{g}(L)-g(L)\right\}\left\{\hat{m}(Z)-m(Z)\right\}\right)\\
&&+E\left[\left\{1-\frac{A}{{g}(L)}\right\}\left\{\hat{Q}(L)-Q(L)\right\}\left\{\hat{m}(Z)-m(Z)\right\}\right]\\
&&+E\left(\left[\frac{A}{\overline{g}^3(L)}\left\{Y-\overline{Q}(L)\right\}\right]\left\{\hat{g}(L)-g(L)\right\}^2\left\{\hat{m}(Z)-m(Z)\right\}\right)\\
&&+E\left(\frac{A}{\overline{g}^2(L)}\left\{\hat{g}(L)-g(L)\right\}\left\{\hat{Q}(L)-Q(L)\right\}\left\{\hat{m}(Z)-m(Z)\right\}\right)\\
&=&E\left(\left\{{Q}(L)-m(Z)\right\}\left\{\hat{m}(Z)-m(Z)\right\}\right)\\
&&+E\left(\frac{g(L)}{\overline{g}^3(L)}\left\{Q(L)-\overline{Q}(L)\right\}\left\{\hat{g}(L)-g(L)\right\}^2\left\{\hat{m}(Z)-m(Z)\right\}\right)\\
&&+E\left(\frac{g(L)}{\overline{g}^2(L)}\left\{\hat{g}(L)-g(L)\right\}\left\{\hat{Q}(L)-Q(L)\right\}\left\{\hat{m}(Z)-m(Z)\right\}\right).
\end{eqnarray*}
for $\overline{g}(L)=tg(L)+(1-t)\tilde{g}(L)$ for some $t\in [0,1]$ and $\tilde{g}(L)$ an element of a vector space large enough to contain $\hat{g}(L)$ and, likewise, $\overline{Q}(L)=tQ(L)+(1-t)\tilde{Q}(L)$ for $\tilde{Q}(L)$ an element of a vector space large enough to contain $\hat{Q}(L)$.
It follows that
\begin{eqnarray*}
E\left[\left\{\hat{m}(Z)-m(Z)\right\}^2\right]&=&{\cal L}(\hat{m},\hat{\eta})-{\cal L}(m,\hat{\eta})\\&&
+2E\left(\left\{{Q}(L)-m(Z)\right\}\left\{\hat{m}(Z)-m(Z)\right\}\right)\\
&&+2E\left(\frac{g(L)}{\overline{g}^3(L)}\left\{Q(L)-\overline{Q}(L)\right\}\left\{\hat{g}(L)-g(L)\right\}^2\left\{\hat{m}(Z)-m(Z)\right\}\right)\\
&&+2E\left(\frac{g(L)}{\overline{g}^2(L)}\left\{\hat{g}(L)-g(L)\right\}\left\{\hat{Q}(L)-Q(L)\right\}\left\{\hat{m}(Z)-m(Z)\right\}\right).
\end{eqnarray*}
By the Cauchy-Schwarz inequality, this can be upper bounded by
\begin{eqnarray*}
E\left[\left\{\hat{m}(Z)-m(Z)\right\}^2\right]&\leq &{\cal L}(\hat{m},\hat{\eta})-{\cal L}(m,\hat{\eta})\\&&
+2E\left(\left\{{Q}(L)-m(Z)\right\}\left\{\hat{m}(Z)-m(Z)\right\}\right)\\
&&+2E\left(\frac{g^2(L)}{\overline{g}^6(L)}\left\{Q(L)-\overline{Q}(L)\right\}^2\left\{\hat{g}(L)-g(L)\right\}^4\right)^{1/2}E\left[\left\{\hat{m}(Z)-m(Z)\right\}^2\right]^{1/2}\\
&&+2E\left(\frac{g^2(L)}{\overline{g}^4(L)}\left\{\hat{g}(L)-g(L)\right\}^2\left\{\hat{Q}(L)-Q(L)\right\}^2\right)^{1/2}E\left[\left\{\hat{m}(Z)-m(Z)\right\}^2\right]^{1/2}.
\end{eqnarray*}
Finally, it follows from the AM-GM inequality that for arbitrary $\delta_1,\delta_2>0$:
\begin{eqnarray*}
E\left[\left\{\hat{m}(Z)-m(Z)\right\}^2\right]&\leq &{\cal L}(\hat{m},\hat{\eta})-{\cal L}(m,\hat{\eta})\\&&
+2E\left(\left\{{Q}(L)-m(Z)\right\}\left\{\hat{m}(Z)-m(Z)\right\}\right)\\
&&+\frac{1}{\delta_1}E\left(\frac{g^2(L)}{\overline{g}^6(L)}\left\{Q(L)-\overline{Q}(L)\right\}^2\left\{\hat{g}(L)-g(L)\right\}^4\right)\\
&&+\frac{1}{\delta_2}E\left(\frac{g^2(L)}{\overline{g}^4(L)}\left\{\hat{g}(L)-g(L)\right\}^2\left\{\hat{Q}(L)-Q(L)\right\}^2\right)\\
&&+(\delta_1+\delta_2)E\left[\left\{\hat{m}(Z)-m(Z)\right\}^2\right],
\end{eqnarray*}
where the second term is non-negative by virtue of minimization (and exactly zero when the function class considered for $m(.)$ contains $E\left\{Q(L)|Z\right\}$).
We conclude that 
\begin{eqnarray*}
E\left[\left\{\hat{m}(Z)-m(Z)\right\}^2\right]&\leq& (1-\delta_1-\delta_2)^{-1}\left[\sup_{\eta} |{\cal L}(m,{\eta})-{\cal L}(\hat{m},{\eta})|\right.
%\\&&\left.+2E\left(\left\{{Q}(L)-m(Z)\right\}\left\{\hat{m}(Z)-m(Z)\right\}\right)\right.
\\
&&\left.+\frac{1}{\delta_1}E\left(\frac{g^2(L)}{\overline{g}^6(L)}\left\{Q(L)-\overline{Q}(L)\right\}^2\left\{\hat{g}(L)-g(L)\right\}^4\right)\right.\\
&&\left.+\frac{1}{\delta_2}E\left(\frac{g^2(L)}{\overline{g}^4(L)}\left\{\hat{g}(L)-g(L)\right\}^2\left\{\hat{Q}(L)-Q(L)\right\}^2\right)\right],
\end{eqnarray*}
where the error $\left\{\hat{g}(L)-g(L)\right\}^4$ and the product of the errors $\left\{\hat{g}(L)-g(L)\right\}^2$ and $\left\{\hat{Q}(L)-Q(L)\right\}^2$ in the last terms make the meta-learner less sensitive to slow convergence of the nuisance parameter estimators, provided that the positivity assumption holds (i.e., that $P(A=1|L)>0$ w.p.1).

\section*{Appendix B}

In this Appendix, we develop oracle bounds for the targeted meta-learner which minimizes the loss
\[{\cal L}(m,\eta)\equiv E\left[A\left\{Y-m(Z)\right\}^2+(1-A)\left\{Q(L)-m(Z)\right\}^2\right],\]
for $\eta\equiv (g,Q)$. In particular, we will use that, by the targeting step, the above loss is sufficiently close to the orthogonal loss considered in Appendix A to deliver favourable oracle bounds.

A Taylor series expansion using directional derivatives shows that
\begin{eqnarray*}
{\cal L}(\hat{m},\hat{\eta})&=&{\cal L}(m,\hat{\eta})
-2E\left[\left\{AY+(1-A)\hat{Q}(L)-m(Z)\right\}\left\{\hat{m}(Z)-m(Z)\right\}\right]\\
&&+E\left[\left\{\hat{m}(Z)-m(Z)\right\}^2\right]\\
&=&{\cal L}(m,\hat{\eta})
-2E\left(\left[\frac{A}{\hat{g}(L)}\left\{Y-\hat{Q}(L)\right\}+\hat{Q}(L)-m(Z)\right]\left\{\hat{m}(Z)-m(Z)\right\}\right)\\
&&+2E\left[\frac{A\left\{1-\hat{g}(L)\right\}}{\hat{g}(L)}\left\{Y-\hat{Q}(L)\right\}\left\{\hat{m}(Z)-m(Z)\right\}\right]\\
&&+E\left[\left\{\hat{m}(Z)-m(Z)\right\}^2\right]
\end{eqnarray*}
from which 
\begin{eqnarray*}
E\left[\left\{\hat{m}(Z)-m(Z)\right\}^2\right]&=&{\cal L}(\hat{m},\hat{\eta})-{\cal L}(m,\hat{\eta})\\&&
+2E\left(\left[\frac{A}{\hat{g}(L)}\left\{Y-\hat{Q}(L)\right\}+\hat{Q}(L)-m(Z)\right]\left\{\hat{m}(Z)-m(Z)\right\}\right)\\
&&-2E\left[\frac{A\left\{1-\hat{g}(L)\right\}}{\hat{g}(L)}\left\{Y-\hat{Q}(L)\right\}\left\{\hat{m}(Z)-m(Z)\right\}\right].
\end{eqnarray*}
By the result of Appendix A (which resulted in identical expressions apart from the last term), 
it follows that 
\begin{eqnarray*}
E\left[\left\{\hat{m}(Z)-m(Z)\right\}^2\right]&\leq& (1-\delta_1-\delta_2)^{-1}\left[\sup_{\eta} |{\cal L}(m,{\eta})-{\cal L}(\hat{m},{\eta})|\right.
%\\&&\left.+2E\left(\left\{{Q}(L)-m(Z)\right\}\left\{\hat{m}(Z)-m(Z)\right\}\right)\right.
\\
&&\left.+\frac{1}{\delta_1}E\left(\frac{g^2(L)}{\overline{g}^6(L)}\left\{Q(L)-\overline{Q}(L)\right\}^2\left\{\hat{g}(L)-g(L)\right\}^4\right)\right.\\
&&\left.+\frac{1}{\delta_2}E\left(\frac{g^2(L)}{\overline{g}^4(L)}\left\{\hat{g}(L)-g(L)\right\}^2\left\{\hat{Q}(L)-Q(L)\right\}^2\right)\right.\\
&&\left.-2E\left[\frac{A\left\{1-\hat{g}(L)\right\}}{\hat{g}(L)}\left\{Y-\hat{Q}(L)\right\}\left\{\hat{m}(Z)-m(Z)\right\}\right]\right].
\end{eqnarray*}
%When $\hat{\eta}$ is calculated on a separate sample of data (as in the proposed cross-fitting procedure) the last term can be rewritten as
%\[-2E\left[\frac{A\left\{1-\hat{g}(L)\right\}}{\hat{g}(L)}\left\{Q(L)-\hat{Q}(L)\right\}\left\{\hat{m}(Z)-m(Z)\right\}\right].\]
%\[-\frac{2}{n}\sum_{i=1}^n\left[\frac{A_i\left\{1-\hat{g}(L_i)\right\}}{\hat{g}(L_i)}\left\{Y_i-\hat{Q}(L_i)\right\}\left\{\hat{m}(Z_i)-m(Z_i)\right\}\right]+o_p(n^{-1/2}).\]
%where the product of the errors $\left\{\hat{g}(L)-g(L)\right\}^2$ and $\left\{\hat{Q}(L)-Q(L)\right\}^2$ in the second term makes the meta-learner less sensitive to slow convergence of the nuisance parameter estimators, provided that the positivity assumption holds (i.e., that $P(A=1|L)>0$ w.p.1).

To bound the last term in the above expression, we will first show that the expectation in that term is close to
\[\frac{1}{n}\sum_{i=1}^n\left[\frac{A_i\left\{1-\hat{g}(L_i)\right\}}{\hat{g}(L_i)}\left\{Y_i-\hat{Q}(L_i)\right\}\left\{\hat{m}(Z_i)-m(Z_i)\right\}\right],\]
which the targeting step in the proposed procedure directly aims to shrink. We will next study the order of magnitude of the latter term.

First, by Markov's inequality, we have that the probability for 
\begin{eqnarray}
&&\frac{1}{\sqrt{n}}\sum_{i=1}^n\left[\frac{A_i\left\{1-\hat{g}(L_i)\right\}}{\hat{g}(L_i)}\left\{Y_i-\hat{Q}(L_i)\right\}\left\{\hat{m}(Z_i)-m(Z_i)\right\}\right]\nonumber\\
&&-E\left[\frac{A\left\{1-\hat{g}(L)\right\}}{\hat{g}(L)}\left\{Y-\hat{Q}(L)\right\}\left\{\hat{m}(Z)-m(Z)\right\}\right]\label{markov}
\end{eqnarray}
to exceed some constant $\epsilon>0$ in absolute value is upper bounded by 
\[\frac{1}{\epsilon^2}E\left[\frac{A\left\{1-\hat{g}(L)\right\}^2}{\hat{g}^2(L)}\left\{Y-\hat{Q}(L)\right\}^2\left\{\hat{m}(Z)-m(Z)\right\}^2\right]\leq \frac{M^2}{\sigma^2\epsilon^2}E\left[\left\{\hat{m}(Z)-m(Z)\right\}^2\right]\]
if $|Y-\hat{Q}(L)|\leq M>0$ with probability 1, as we will assume; note that this is guaranteed to hold for dichotomous outcomes, whose analysis motivated this work. It thus follows that 1 over root-$n$ times (\ref{markov}) is of the order $O_p(n^{-1/2})E\left[\left\{\hat{m}(Z)-m(Z)\right\}^2\right]$ (and  thus a lower order term).
%Further, since we focus on dichotomous exposures and outcomes, so that $\hat{Q}(L),m(Z)$ and $\hat{m}(Z)$ are bounded, then under a positivity assumption that $g(L)>\sigma>0$ and restricting $\hat{g}(L)>\sigma$, we have that $C(\hat{\eta})\equiv \frac{A\left\{1-\hat{g}(L)\right\}}{\hat{g}(L)}\left\{Y-\hat{Q}(L)\right\}$ and $\left\{\hat{m}(Z)-m(Z)\right\}$ are bounded, with $C(\hat{\eta})\left\{\hat{m}(Z)-m(Z)\right\}\in [1-\sigma^{-1},\sigma^{-1}-1]$. The Rademacher complexity of the set of functions $C_i(\eta),i=1,...,n$ with $\eta$ obeying the above positivity condition is upper bounded by $\sigma^{-1}$. Since furthermore $||C(\eta)||_{\infty}\leq \sigma^{-1}$, it follows from Theorem 4.10 in Wainwright that for arbitrary $\delta>0$
%\[\sup_{\eta}|\frac{1}{n}\sum_{i=1}^n C_i(\eta)\left\{\hat{m}(Z_i)-m(Z_i)\right\}-E\left[C(\eta)\left\{\hat{m}(Z)-m(Z)\right\}\right]\right\}|\] is upper bounded by $2\sigma^{-1}+\delta$ with probability at least $1-\exp(-n\delta^2\sigma^2/2)$.

Next, denote
\[C_i\equiv A_i\frac{1-\hat{g}(L_i)}{\hat{g}(L_i)}\left\{Y_i-\hat{Q}(L_i)\right\},\]
then it remains to study the order of magnitude of
\[\frac{1}{n}\sum_{i=1}^n C_i\left\{\hat{m}(Z_i)-m(Z_i)\right\}.\]
In the proposed procedure based on penalized sieve estimators, 
we %have that 
%$\hat{Q}(L_i)=\hat{Q}^{(0)}(L_i)+\sum_{j=1}^{J_n}\hat{\beta}_jb_j(Z_i)$ and 
restrict the optimization procedure to functions of the form $m(Z_i)=\sum_{j=1}^{\infty}\gamma_jb_j(Z_i)$. Then
\begin{eqnarray*}
\frac{1}{n}\sum_{i=1}^n C_i\left\{\hat{m}(Z_i)-m(Z_i)\right\}
&=&
\frac{1}{n}\sum_{i=1}^n\sum_{j=1}^{J_n} C_ib_j(Z_i)(\hat{\gamma}_j-\gamma_j)-\frac{1}{n}\sum_{i=1}^n\sum_{j=J_n+1}^{\infty} C_ib_j(Z_i)\gamma_j,
\end{eqnarray*}
since $\hat{\gamma}_j=0$ when $j>J_n$.
The penalization procedure used for the targeting step ensures (by H\"older's inequality) that the first term
\begin{eqnarray*}
\left\lVert\frac{1}{n}\sum_{i=1}^n\sum_{j=1}^{J_n} \epsilon_ib_j(Z_i)(\hat{\gamma}_j-\gamma_j)\right\rVert
&\leq& \left\lVert\frac{1}{n}\sum_{i=1}^n\sum_{j=1}^{J_n} \epsilon_ib_j(Z_i)\right\rVert_{\infty} ||\hat{\gamma}_j-\gamma_j||_1\leq \lambda ||\hat{\gamma}_j-\gamma_j||_1
\end{eqnarray*}
Here, the last inequality follows from the Karush-Kuhn-Tucker conditions, with $\lambda$ the penalty term, which is of the order $O_p\left(\sqrt{\log{J_n}/n}\right)$ with $J_n=C(D)d^{D'}n^{1/3}\log^{D'-1}{(n)}$ (see Theorem 5.1 in \cite{zhang2022regression}) under the ellipsoid-type condition which restricts $m(Z)$ to satisfy
\[\sum_{j=1}^{\infty} \left(\frac{j}{\max{(\log^{D-1}{j},1)}}\right)^{2}\gamma_j^2\leq Q\]
for some constant $Q$. Further, from Corollary D.6 in \cite{zhang2022regression} with $s=1$ (in line with the above ellipsoid-type condition), 
\[||\hat{\gamma}_j-\gamma_j||_1
=O_p\left(\sqrt{\frac{\log{J_n}}{n}}n^{\frac{1}{3}}\log^{\frac{2(D-1)}{3}}{(n)}\right)\]
We conclude that the first term is
\[
O_p\left(\log{J_n}n^{-2/3}\log^{\frac{2(D-1)}{3}}{(n)}\right)\]

Further, we have that 
\[\frac{1}{n}\sum_{i=1}^n\sum_{j=J_n+1}^{\infty} C_ib_j(Z_i)\gamma_j=\frac{1}{n}\sum_{i=1}^n \epsilon_i \sum_{j=J_n+1}^{\infty}  b_j(Z_i)\gamma_j+\frac{1}{n}\sum_{i=1}^n\sum_{j=J_n+1}^{\infty} C^*_i b_j(Z_i)\gamma_j,\]
where 
\[\epsilon_i\equiv A_i\frac{1-\hat{g}(L_i)}{\hat{g}(L_i)}\left\{Y_i-{Q}(L_i)\right\},\]
and 
\[C_i^*\equiv A_i\frac{1-\hat{g}(L_i)}{\hat{g}(L_i)}\left\{Q(L_i)-\hat{Q}(L_i)\right\}.\]
Here, the first term has mean zero (namely, it has mean zero given $A$ and $L$ since $\hat{g}(L_i)$ is a functional of only the data in $A$ and $L$) and variance
\begin{eqnarray*}
\frac{1}{n}E\left[\mbox{\rm Var}(\epsilon|Z)\left\{\sum_{j=J_n+1}^{\infty}  b_j(Z_i)\gamma_j\right\}^2\right]
&\leq & \frac{1}{n}\sup_z \mbox{\rm Var}(\epsilon|z)f_Z(z)
\int \left\{\sum_{j=J_n+1}^{\infty} b_j(z)\gamma_j\right\}^2dz\\
&\leq & \frac{1}{n}\sup_z \mbox{\rm Var}(\epsilon|z)f(z)
\sum_{j=J_n+1}^{\infty} \gamma^2_j,
\end{eqnarray*}
where the last step follows from the chosen basis being orthonormal, and where we assume that the conditional variance $\mbox{\rm Var}(\epsilon|.)$ and the density $f_Z(.)$ are bounded (which is a plausible assumption when there are no positivity violations).
Further, by the approximation results for sieves in lemma C.5 and C.6 of \cite{zhang2022regression} (see in particular the similar derivation leading to equation (60) in \cite{zhang2022regression}), we have that $\sum_{j=J_n+1}^{\infty} \gamma^2_j$ is of the order 
$\left(\log^{D-1}{n}/n\right)^{\frac{2}{3}}$ (or less).
We conclude (by calling on Markov's inequality) % see start of proof of Theorem 2 in Kennedy's paper on continuous exposures
that the first term is $O_p\left(n^{-1/2}(\log^{D-1}{n}/n)^{\frac{1}{3}}\right)=O_p\left(n^{-5/6}\log^{(D-1)/3}{n}\right)$.

We next bound the second term.
First, note that, by
a similar reasoning as in the previous paragraph, its variance
is readily shown to be upper bounded by a term of the order 
$n^{-1}\left(\log^{D-1}{n}/n\right)^{\frac{2}{3}}$.
We therefore conclude that the second term equals
\[E\left\{C^* \sum_{j=J_n+1}^{\infty} b_j(Z)\gamma_j\right\}
+O_p\left(n^{-5/6}\log^{(D-1)/3}{n}\right)\]
Using the Cauchy-Schwarz inequality, we further have that
\begin{eqnarray*}
\left\lVert E\left\{C^* \sum_{j=J_n+1}^{\infty} b_j(Z)\gamma_j\right\}\right\rVert
&\leq& E\left(C^{*2}\right)^{1/2}E\left\{\left(\sum_{j=J_n+1}^{\infty}  b_j(Z)\gamma_j\right)^2\right\}^{1/2}.
\end{eqnarray*}
By a similar reasoning as in the previous paragraph, 
\[E\left\{\left(\sum_{j=J_n+1}^{\infty}  b_j(Z)\gamma_j\right)^2\right\}\]
is readily shown to be upper bounded by a term of the order 
$\left(\log^{D-1}{n}/n\right)^{\frac{2}{3}}$. We conclude that \[E\left\{C^* \sum_{j=J_n+1}^{\infty} b_j(Z)\gamma_j\right\}\]
is of the order $O(n^{-b-1/3}\log^{(D-1)/3}{n})$ for some $b\geq 1/2$.

Putting it all together, we conclude that 
\begin{eqnarray*}\frac{1}{n}\sum_{i=1}^n\sum_{j=J_n+1}^{\infty} C_ib_j(Z_i)\gamma_j&=&
O_p\left(\log{J_n}n^{-2/3}\log^{\frac{2(D-1)}{3}}{n}\right)
+ O_p\left(n^{-5/6}\log^{(D-1)/3}{n}\right)\\&&
+O_p(n^{-b-1/3}\log^{(D-1)/3}{n})\end{eqnarray*}

\section*{Appendix D}

\begin{figure}[ht!]
\begin{center}
\includegraphics[width=\textwidth]{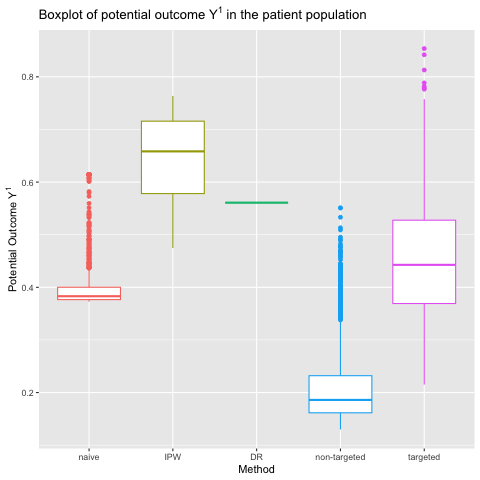}
\end{center}
\caption{Boxplot of the $7$-day ICU mortality under initiation of RRT within $24$h from the stage $2$ AKI diagnosis in the stage $\geq2$ AKI patient population, i.e., the potential outcome $Y^1$, conditional on the values of age on admission, gender, serum potassium, arterial pH, fluid intake and fluid output computed in the whole patient population using "naive" approach ("naive"), IPW-learner ("IPW"), DR-learner ("DR"), proposed targeted learner, i.e. i-Learner ("targeted"), and its non-targeted equivalent ("non-targeted"), using 20 basis functions.} 
\end{figure}

\begin{figure}[ht!]
\begin{center}
\includegraphics[width=\textwidth]{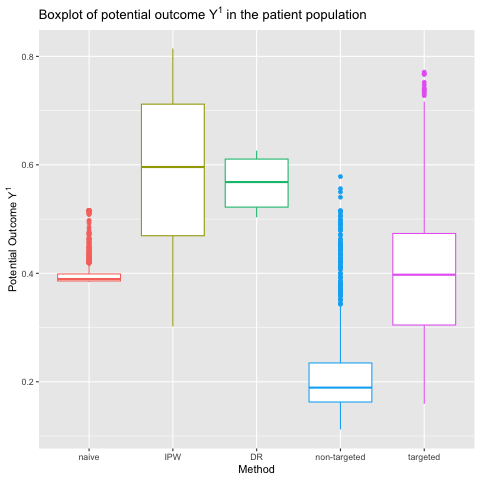}
\end{center}
\caption{Boxplot of the $7$-day ICU mortality under initiation of RRT within $24$h from the stage $2$ AKI diagnosis in the stage $\geq2$ AKI patient population, i.e., the potential outcome $Y^1$, conditional on the values of age on admission, gender, serum potassium, arterial pH, fluid intake and fluid output computed in the whole patient population using "naive" approach ("naive"), IPW-learner ("IPW"), DR-learner ("DR"), proposed targeted learner, i.e. i-Learner ("targeted"), and its non-targeted equivalent ("non-targeted"), using 50 basis functions.} 
\end{figure}

\begin{figure}[ht!]
\begin{center}
\includegraphics[width=\textwidth]{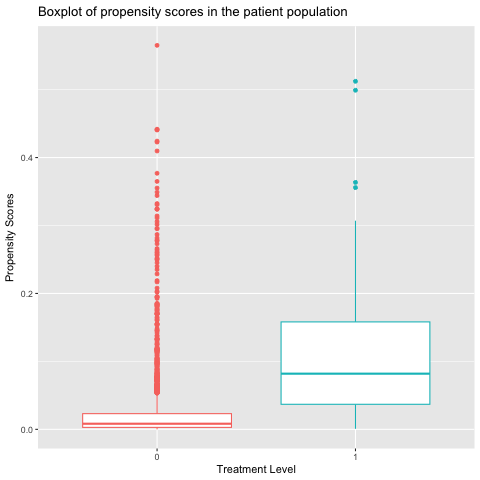}
\end{center}
\caption{Boxplot of the propensity scores in both treatment groups. Treatment level "1" denotes "RRT initiation" and treatment level "0" denotes "no RRT initiation".}
\end{figure}

\end{document}